\input{aipcheck}

\documentclass[
    ,final            
  ]
  {aipproc}

\layoutstyle{6x9}

\begin{document}

\title{Gamma-ray emission from Wolf-Rayet stars interacting with AGN jets}


\author{A.T. Araudo}{
  address={Centro de Radioastronom\'{\i}a y Astrof\'{\i}sica, 
Universidad Nacional Aut\'onoma de M\'exico, A.P. 3-72 (Xangari), 
58089 Morelia, Michoac\'an, M\'exico}
}
\author{V. Bosch-Ramon}{
  address={Universitat de Barcelona.
Departament d'Astronomia i Meteorologia
Marti i Franques 1, 7th floor
ES 08028 Barcelona
Spain}
}
\author{G.E. Romero}{
  address={Instituto Argentino de
Radioastronom\'{\i}a, C.C.5, (1894) Villa Elisa, Buenos Aires.\\
Facultad de Ciencias Astron\'omicas y Geof\'{\i}sicas,
Universidad Nacional de La Plata, Paseo del Bosque, 1900 La Plata,
Argentina}
}

\begin{abstract}
Dense populations of stars surround the nuclear
regions of galaxies.
In this work, we study the interaction of a WR star with
relativistic jets in active galactic nuclei. A bow-shaped double-shock 
structure will form as a
consequence of the interaction of the jet and the  wind of the star.
Particles can be accelerated up to relativistic
energies in these shocks and emit high-energy radiation. We compute
the produced $\gamma$-ray emission obtaining that this 
radiation may be significant. This emission is expected 
to be particularly relevant for nearby non-blazar sources. 
\end{abstract}

\maketitle


\section{Introduction}
Active galactic nuclei (AGNs) consist of a supermassive
black hole (SMBH) surrounded by an accretion disc in the center of a galaxy. 
Sometimes these objects 
present radio emitting jets originated close to the SMBH.
Jets of AGN are relativistic ($v_{\rm j} \sim c$), with  macroscopic
Lorentz factors $\Gamma \sim 5-10$, and density 
$\rho_{\rm j} = L_{\rm j}/[(\Gamma -1)c^2\sigma_{\rm j}v_j]$,
where $L_{\rm j}$ and
$\sigma_{\rm j} = \pi R_{\rm j}^2$ are the jet kinetic luminosity and
section, respectively, and $R_{\rm j}$ its radius. 
According to the
current taxonomy of AGN, jets from type I Faranoff-Riley galaxies
(FR~I) are low luminous,  with a kinetic luminosity $L_{\rm j}
<10^{44}$~erg~s$^{-1}$,  whereas FR~II jets have $L_{\rm j} >
10^{44}$~erg~s$^{-1}$.

In the nuclear region of AGNs there is matter in the form of
diffuse gas,  clouds, and stars, making jet medium interactions
likely. Different models based on the interaction of 
jets with obstacles from the external medium 
have been proposed in order to explain the
$\gamma$-ray emission produced in misaligned AGN jets [1,2].  
In the present contribution we study a new scenario: the interaction of 
Wolf-Rayet (WR) stars with the jets.

\section{Jet-star interaction}

We consider that a WR star  with mass loss rate  
$\dot M_{\rm w} = 10^{-4}$~M$_{\odot}$~yr$^{-1}$ and    
terminal wind velocity $v_{\infty} = 3000$~km~s$^{-1}$  
penetrates the jet at $z_{\rm int} = 5\times10^{-4}$~pc, that correspond 
to a value of
$10$ times the base of the jet that emanates from a SMBH of mass 
$10^{7}$~M$_{\odot}$.
When the jet interacts with the star a double bow shock is formed around it,
as is shown in Figure~1. 
The location of the stagnation point is at a distance $R_{\rm sp}$
from the stellar surface, where the wind and jet ram pressures are equal.
From $\rho_{\rm w}\,v_{\infty}^2 = \rho_{\rm j}\,c^2\,\Gamma$, where
$\rho_{\rm w} \sim \dot M_{\star}/(4\pi R_{\rm sp}^2 v_{\rm \infty})$ is 
the wind density, we obtain
\begin{equation}
\frac{R_{\rm sp}}{R_{\rm j}} \sim  
0.1 \left(\frac{\dot M_{\rm w}}{10^{-4}\,{\rm M_{\odot}\,yr^{-1}}}\right)^{1/2}
\left(\frac{v_{\infty}}{3000\,{\rm km\,s^{-1}}}\right)^{1/2}
\left(\frac{L_{\rm j0}}{10^{42}\,{\rm erg\,s^{-1}}}\right)^{-1/2}
\left(\frac{\Gamma -1}{9}\right)^{1/2}.
\end{equation}

\begin{figure}
\includegraphics[height=.3\textheight]{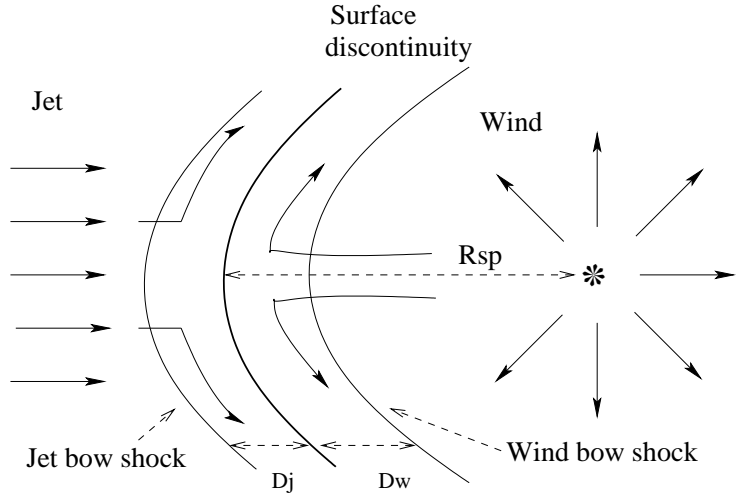}
\caption{Left: Sketch of the scenario considered in the present study. 
}
\end{figure}

\section{Particle acceleration and losses}

Particles can
be accelerated up to relativistic energies in both
the jet  and wind  shocks. Relativistic electrons and protons are
 injected in the downstream regions following a distribution
$Q_{e,p} \propto E_{e,p}^{-2}$. The luminosity of particles accelerated in the
jet and in the wind bow shocks is 
$L_{\rm ntj} \sim 0.1(R_{\rm sp}/R_{\rm j})^2L_{\rm j}$ and 
$L_{\rm ntw} \sim 0.1 L_{\rm w}/4$, respectively,  where 
$L_{\rm w} = \dot M_{\rm w} v_{\infty}^2/2$. 
We estimate the magnetic field in the jet shocked region, $B_{\rm jbs}$, 
assuming that the magnetic energy density is a fraction $\eta_{\rm B}$ 
of the energy density of the jet shocked matter,   
resulting in 
\begin{equation}
B_{\rm jbs}  \sim 17
\left(\frac{\eta_{\rm B}}{0.01}\right)^{1/2}
\left(\frac{L_{\rm j}}{10^{42}\,{\rm erg\,s^{-1}}}\right)^{1/2}
\left(\frac{z}{z_{\rm int}}\right)^{-1}\,\,{\rm G}.
\end{equation}
The main radiative losses that affect the evolution of
$Q_{e}$ are synchrotron radiation and
Inverse Compton (IC) scattering. For the later we have considered 
stellar target photons with an energy $\sim 7.9$~eV and luminosity 
$L_{\star} = 10^{39}$~erg~s$^{-1}$. 
In addition to radiative losses, electrons can escape from the 
emitter by advection or diffusion.
For the wind we assume the  parametrization of the 
magnetic field $B_{\rm w}$ given in [5], with a value
in the stellar surface of about 10~G. 
    
At $z_{\rm int}$, the maximum energy of electrons accelerated in the
jet bow shock is constrained by synchrotron losses, reaching a value
$E_e^{\rm max} \sim 3$~TeV. Inverse Compton  scattering is also an
important channel of electron cooling (in the jet and in the wind) 
as a consequence of the large
value of the energy density of the WR radiation field at $R_{\rm sp}$:
$U_{\rm ph \star}\sim 2.4$~erg~cm$^{-3}$. This process constrains 
the maximum energy of electrons accelerated in the wind bow shock, yielding
a value of $\sim 10$~GeV. The decay of $\pi^0$
produced in $pp$ interactions is not relevant compared with IC
emission (see Fig.~2-\emph{Left}). Bolometric luminosities achieved by 
different radiative processes and jet luminosities
are listed in  Table~\ref{bolometrica}. 
Absorption  of $\gamma$ rays by
stellar photons is important at  photon energies $E_{\rm ph} > 30$~GeV. 
In Fig.~2 (\emph{Right}), the computed spectral energy
distributions (SEDs) for the 
cases of $L_{\rm j} = 10^{42}$, $10^{45}$, and $10^{48}$~erg~s$^{-1}$ are shown.

\begin{table}
\caption{Bolometric luminosities of synchrotron emission 
($L_{\rm syn}^{\rm j,w}$) and IC scattering ($L_{\rm IC}^{\rm j,w}$) in the jet
and in the wind.
Values are given in erg~s$^{-1}$ units.}
\label{bolometrica}
\begin{tabular}{l|lll||l|ll}
\hline
{}&  $L_{\rm j} =10^{42}$ & $L_{\rm j} =10^{45}$ & $L_{\rm j} =10^{48}$&
{}& $L_{\rm j} =10^{42}$ & $L_{\rm j} = 10^{45}$\\
\hline
$L_{\rm syn}^{\rm j}$&$8.6\times10^{38}$&$1.2\times10^{39}$&$3.2\times10^{42}$&
$L_{\rm syn}^{\rm w}$&$1.2\times10^{33}$&$1.0\times10^{36}$\\
$L_{\rm IC}^{\rm j}$&$9.1\times10^{38}$& $3.3\times10^{39}$&$4.4\times10^{40}$&
$L_{\rm IC}^{\rm w}$&$6.5\times10^{37}$& $4.1\times10^{39}$\\
\hline 
\end{tabular}
\end{table}

\begin{figure}
  \includegraphics[height=.35\textheight, angle=-90]{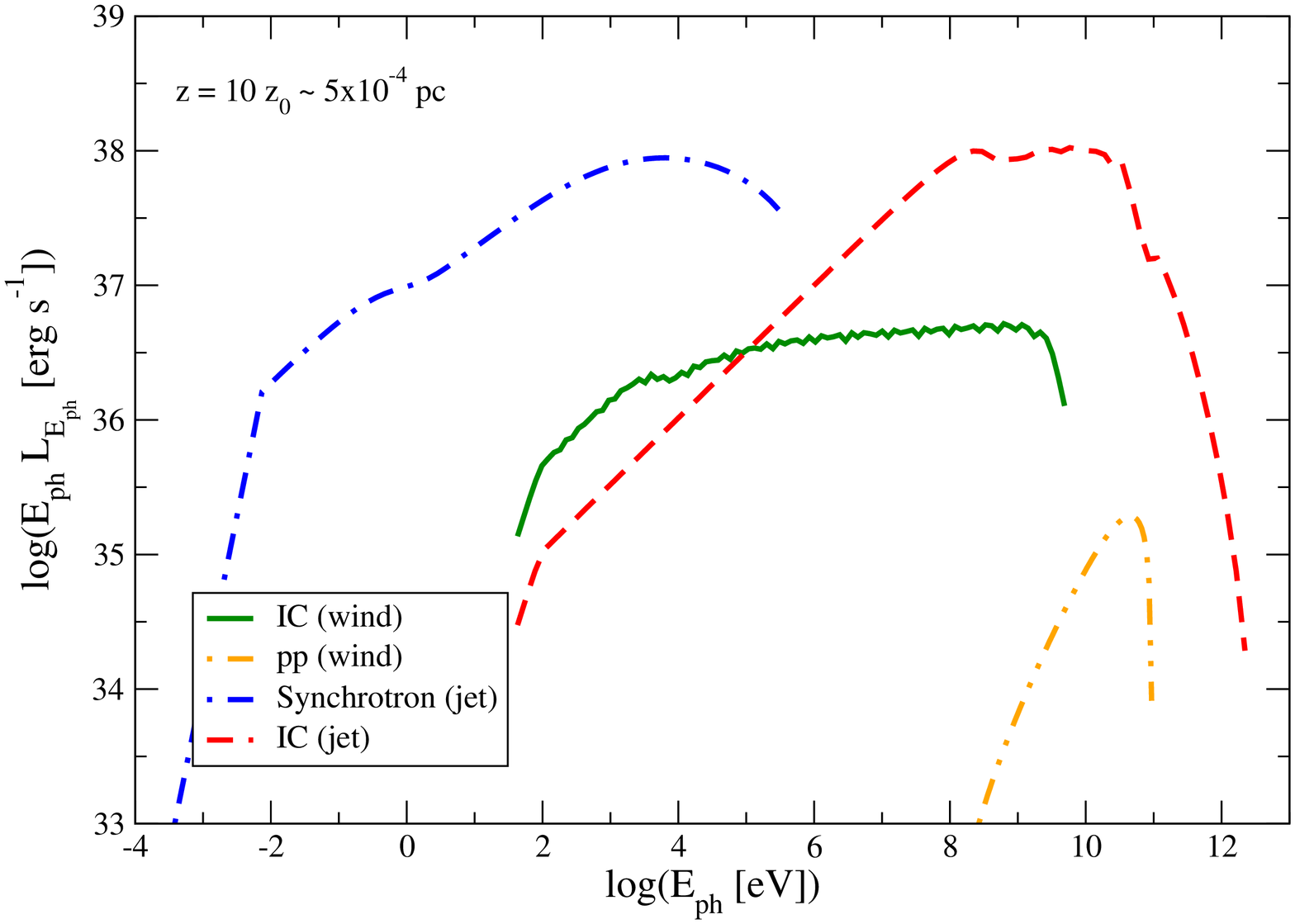}
  \includegraphics[height=.35\textheight, angle=-90]{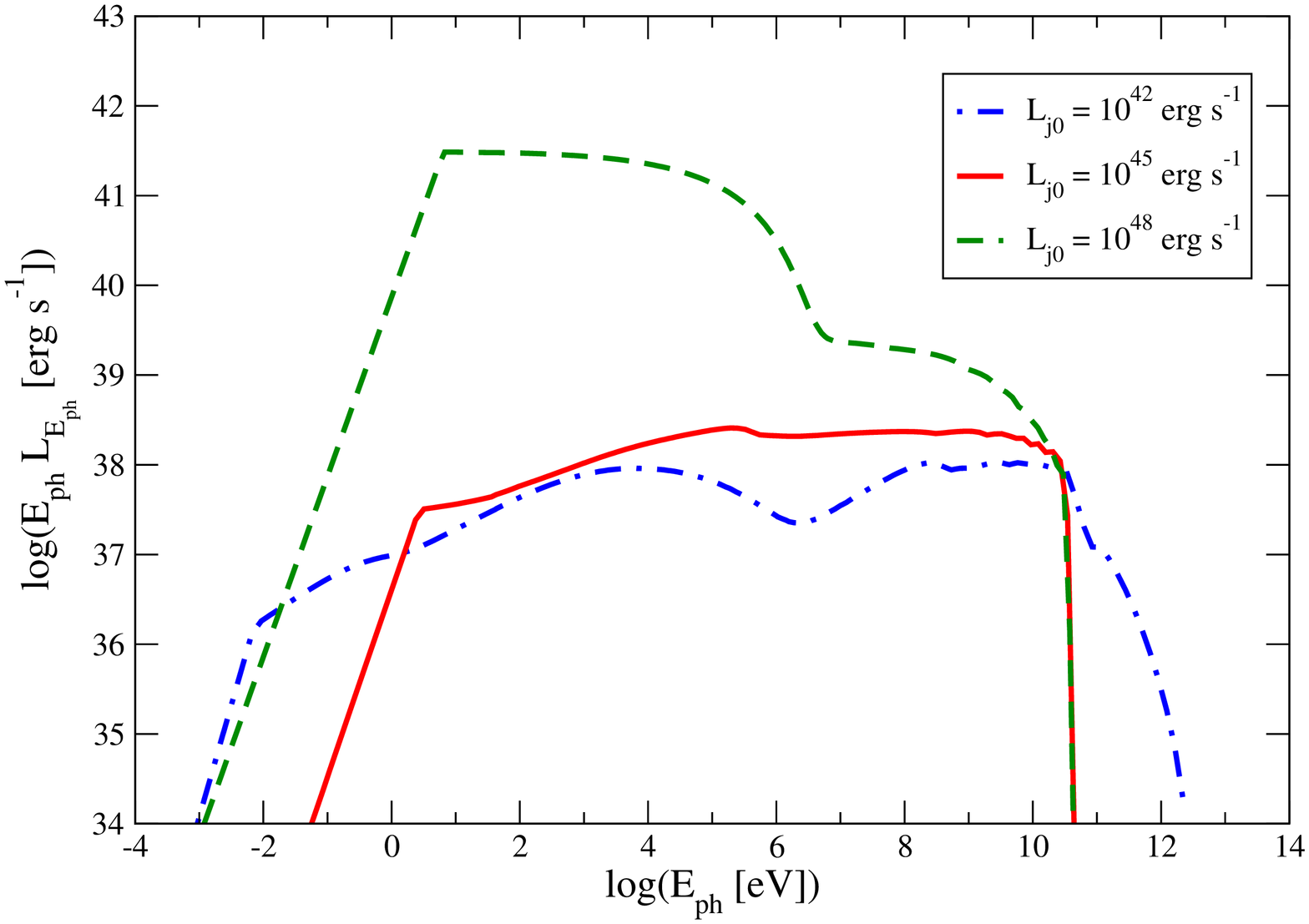}
  \caption{\emph{Left}: 
Synchrotron emission,  IC scattering (in the jet and in the wind), and
$pp$ components for the case of  $L_{\rm j} = 10^{42}$~erg~s$^{-1}$.
\emph{Right}: Spectral energy distributions where all the 
contributions were added.  The main contributions
to the SED are synchrotron radiation 
and IC scattering in the jet. In the case of $L_{\rm j0}= 10^{42}$~erg~s$^{-1}$
IC radiation in the wind is also relevant.}
\end{figure}

\subsection{Gamma-ray emission}

The achieved emission levels in $\gamma$ rays  in the case of 
$L_{\rm j} = 10^{48}$~erg~s$^{-1}$ ($\sim 3\times10^{39}$~erg~s$^{-1}$) could be 
detectable by the \emph{Fermi} satellite
in nearby AGNs.  The radiation
produced by a WR interacting time to time with a jet will be
transient. It is noteworthy that
one or few WR may be permanently present within the jet at
$z> z_{\rm int}$, where radiative cooling is still dominant, adding up to
the contribution of the many-star persistent emission studied in [5]. 
In fact, WR
could be important contributors of their own to the non-thermal output
of misaligned AGN jets.

\section{Discussion}

The interaction of a WR star with the jet can produce significant 
amounts of $\gamma$ rays only if the interaction height is below the $z$
at which advection escape dominates the whole particle
population. Also, $\sigma_{\rm sp}$ should be a significant fraction
of $\sigma_{\rm j}$.  In this context, we have considered the
interaction of a powerful WR star at $z = 5\times10^{-4}$~pc (for which the bow
shock  covers $\sim 1\%$ of the jet section).
The emission produced by IC scattering achieves values as
high as $5\times10^{39}$~erg~s$^{-1}$ in the  \emph{Fermi} range. Such an
event would not last long though, about $R_{\rm j}/v_{\star}\sim
10^7\,(R_{\rm j}/10^{16}\,{\rm cm})\,(10^9\,{\rm cm~s}^{-1}/v_{\star})^{-1}$~s,
where $v_{\star}$ is the velocity of the WR.

Since jet-star emission should be rather isotropic, it would be masked
by jet beamed emission in blazar sources. However, when radio loud AGN
jets do not display significant beaming, these
objects may emit detectable $\gamma$ rays from jet-star interaccions. 
The emission level achieved by the interaction of a WR with a jet 
close to the jet base could be detectable
by \emph{Fermi} only for very nearby sources, like Centaurus~A or
M87. The interaction of a star even more powerful than a WR, like a
Luminous Blue Variable, may provide $R_{\rm sp}\sim R_{\rm j}$, making
available the whole jet luminosity budget for particle acceleration.
After few-year exposure times of \emph{Fermi}, a significant 
signal from close and powerful
sources could be detectable. Their detection can shed light 
on the jet matter composition
as well as on the stellar populations in the vicinity of AGNs.

\section{Acknowledgments}
A.T.A. is very grateful for the
hospitality of the Dublin Institute of Advanced Studies (DIAS) where
this project started. 
This work is supported by CONACyT, Mexico and PAPIIT, UNAM; and by  
PIP 0078/2010 from CONICET
and PICT 848/2007 of Agencia de Promoci\'on Cient\'{\i}fica y T\'ecnica, 
Argentina.
G.E.R. and V.B-R. acknowledge support by the Ministerio de 
Ciencia e Innovaci\'on  (Spain) under grant AYA 2010-21782-C03-01. 
V.B-R. also acknowledges support by the Ministerio de 
Ciencia e Innovaci\'on (Spain) under grant FPA2010-22056-C06-02.

\bibliographystyle{aipproc}   

\end{document}